# Legal Concerns and Challenges in Cloud Computing

Sundar Krishnan [1] and Lei Chen [2]

[1] Department of Computer Science, Sam Houston State University, Huntsville, Texas, USA, sxk030@shsu.edu
[2] Department of Computer Science, Sam Houston State University, Huntsville, Texas, USA, chen@shsu.edu

*Abstract*— Legal issues have risen with the changing landscape of computing, especially when the service, data and infrastructure is not owned by the user. With the Cloud, the question arises as to who is in the "possession" of the data. The Cloud provider can be considered as a legal custodian, owner or possessor of the data thereby causing complexities in legal matters around trademark infringement, privacy of users and their data, abuse and security. By introducing Cloud design focusing on privacy, legal as a service on a Cloud and service provider accountability, users can expect the service providers to be accountable for privacy and data in addition to their regular SLAs.

## I. INTRODUCTION

Cloud computing in layman's terms can be defined as a network of virtual computers hosted outside our firewalls. Cloud computing caters to the demands of Information Technology in terms of increasing capacity and features with decreasing costs. Thus, accessing the Cloud comes with a cost known as the subscription fee per service, also known as service models. Users (post authentication) can access the Cloud using browsers on their tablets, desktops and laptops.

This paper begins with discussing the two most popular issues of user-privacy and data-privacy challenges on the Cloud. The paper also discusses a few solutions to these issues. It concludes with a discussion on Cloud designs, Accountability and LaaS (Legal as a Service) as options that can be provided by the Service Providers to the data users via the SLAs.

### A. *User privacy in the Cloud*

Privacy is a very important consideration in the Cloud computing world since actual or perceived privacy weaknesses will impact compliance, data security and user trust thereby giving rise to legal complications [1][5]. Unfortunately, legal rights and regulatory authority for the protection of the user privacy in the Cloud computing world is not well defined [9].

Access and storage of user's movements and behavioral information while on the Cloud has a huge market for the data mining and advertising companies. Information such as viewing habits within the Cloud generates huge statistics that has a market. As user's movement on the Cloud is tracked and stored, this data is of very high interest to many companies. They study the movements to in turn spam the user with products that they may want in the near future. For example, a user looking for a mobile phone may also need a service plan. Thus, smart programs working in the background pop-up messages prompting the user with various service plans from different cell phone carriers.

Companies would like to know the patterns of user movement within the Cloud thereby better enabling them to setup their products so that the user can be attracted to them. Data-mining companies also take interest in how we search and apply for jobs and where we get our news, to how we find friends. Often the service providers do not let the user know that their presence on the Cloud could lead to collection and marketing of such statistics. Also, many users do not feel the importance of such risks until it's too late.

In summary, since Cloud computing is still evolving, service providers often change their policies, SLAs and Operating *ToS* and are not obligated to inform the users. In many cases, they do notify the users on their websites, but, user ignorance steers them to ignore such information updates. Despite user ignorance of the challenges of using the Cloud, users are embracing the Cloud and taking advantage of its benefits like low costs, reliability, security and simplicity. Due to its sudden surge in popularity, Cloud computing may find itself a prey to security, privacy and legal issues.

## II. BACKGROUND

### A. *Data Privacy - Laws and Acts*

When users place their data and applications on the Cloud servers, they lose the ability to maintain complete control of that information. The critical and sometimes sensitive information that was once safely stored on personal computers now resides on the servers of online companies. Such data security concerns prevent companies and users from taking advantage of the Cloud. Security concerns can be of 3 types: Traditional Security, Availability, and Third-party data control [7].

Traditional security concerns involve Virtual Machine level attacks, cross-site scripting, and phishing of a Cloud provider, computer and network intrusions, attacks or hacking. Availability concerns center on critical applications up-time, single point of failure and assurance of computational integrity. Third-party control of data concerns are about legal implications of data location, loss, data-audit, contractual obligations, data lock-ins etc.

*1)* Electronic Communications Privacy Act*:* Under this Act, data stored in the Cloud may be subject to a lesser standard for law enforcement to gain access to it than if the data were stored on a personal computer. Moreover, the ToS and SLAs for Cloud services often makes it clear that they will preserve and disclose information to law enforcement when served with a legal process. Thus, data privacy issues exist such as appropriate collection of data, appropriate data use, data disclosure, safe data storage, retention of data, data access and ways of keeping the user informed about how these issues are handled and impact them.

*2)* Stored Communications Act: Cloud computing allows users to store and access their files and data away from their personal machines. The Cloud is seen as a single application or a device that can be accessed from various computing devices. Although users might expect that their data stored on the Cloud is private, in reality they do not enjoy a lot of privacy. By passing the Stored Communications Act (SCA) [19], Congress hoped to encourage development and use of new and emerging methods of communications by protecting citizen's privacy rights. The SCA limits the government's ability to compel Internet service providers to disclose information stored with them. The Act defines service providers as Electronic Communications Services (ECS) and Remote computing Services (RCS). The level of privacy protection afforded by a stored communication differs based on which category the service provider falls in and sometimes, for how long the communication was stored.

In summary, as the current Stored Communications Act is outdated and complicated, the courts have interpreted it in an inconsistent and unclear manner [14]. The extent to which protections under the SCA apply is an open question and depends on the courts applying the reasoning of *Theofel (Theofel v. Farey-Jones* 359 F.3d 1066 (9th Cir. 2004) [18]).

*3)* Federal Information Security Management Act (FISMA): This Act was enacted in 2002 to recognize the importance of information security to the economic and national security interests of the United States. It provides a uniform regime to address levels of risk that may arise from domestic and international sources. It requires federal agencies to create and implement programs to review information security and report the results to the Office of management and Budget (OMB).

With the Federal agencies taking to the Cloud to reduce costs, security and data privacy concerns are primary reasons for not migrating their systems into the Cloud. Also, they are concerned about losing control and thus want visibility into the Cloud's security incidents and risk management.

The General Services Administration (GSA) and Office of Management and Budget (OMB) have focused on security and data privacy as top priorities to facilitate Cloud adoption through the Federal Cloud Computing Initiative, GSA's Blanket Purchase Agreement (BPA) for Cloud Infrastructure as a Service (IaaS) and the Federal Risk and Authorization Management Program (FedRAMP), the government-wide program providing a standardized approach to security assessment, authorization and continuous monitoring for Cloud products and services [15].

*4)* Fourth Amendment issues: The Fourth Amendment [14] provides that "The right of the people to be secure in their persons, houses, papers, and effects, against unreasonable searches and seizures, shall not be violated, and no Warrants shall issue, but upon probable cause, supported by Oath or affirmation, and particularly describing the place to be searched, and the persons or things to be seized" [15]. The architecture of the Internet and Cloud is such that the courts are unlikely to apply the Fourth Amendment protections for users.

Since the information is inherently handled and processed by third parties, it may be difficult to separate coding information from protected content. Also, it is unclear as to which machine or set of machines on the Cloud would be considered as the "container" in the warrant.

*5)* The USA PATRIOT ACT: This Act gives FBI access to any business record as long as a court order is issued. This privilege can be used to obtain data from the Cloud without the knowledge of the user. This can increase the mistrust between the users and the governmental agencies and could deter users from using the Cloud services.

*6)* Jurisdictional Issues: In the Cloud design, users can access their data from any location as the data can be stored on distributed virtual servers in data centers spread across many countries. Cloud service providers consider the location of the data centers for purposes like costs, laws, infrastructure and labor. Thus, the question arise "Which country's laws to apply"? User data is stored across many data centers in the world and is dependent on the service provider's agreements with the operators of the data-centers. Legal experts are wary of cases involving a Cloud. Also, if software development is conducted on a Cloud, Copyright issues could arise from country to country depending on which machine was used for which developer [17]. Such scenarios are further complicated when developers are scattered around the world. Thus, the locations of the data centers affect the legal rights of the users.

*7)* HIPPA: Health information service providers who *store* user medical information may not be subject to the privacy protections of the Health Insurance Portability Protection Act. Even when it is clear that user data is protected, Cloud service providers often limit their liability to the user as a condition of providing the Cloud service leaving users with limited recourse should their data be exposed or lost [10].

III. SOLUTIONS TO LEGAL CHALLENGES

A. *Privacy Impact Assessment (PIA)*

A Privacy Impact Assessment (PIA) is a predictive and proactive systematic business process to evaluate possible future effects that a particular activity or task that may have on user's privacy. It focuses on understanding the system thereby identifying and mitigating any adverse privacy impacts. It informs decision makers who must decide whether the task should proceed into the next step and if so; in what form.

Although reactive processes such as privacy issue analysis, privacy audits and privacy law compliance checking can be applied to existing systems, a proactive measure can be well managed if planned well.

Privacy Impact Assessment (PIA) was initially launched by the Information Commissioners Office (ICO) of UK to help organizations access the impact of their Cloud operations on personal privacy. This process was primarily intended for use in the public sector risk management but has been a value to private sector businesses that process personal data. The role of a PIA within the Cloud is to ensure that the risks to personal privacy are mitigated and should be initiated early in the design phase and needs to be revisited in every phase. The output of this process needs to undergo corrective action and the fed back into the next stage in the design process in an iterative manner.

In conclusion, many PIA tools have now been designed and have been well embraced by organizations when dealing with the Cloud. A typical PIA tool contains a set of questions and answers with calibrated weightage [10]. A drawback of this tool is that the organizations or users need to drive its implementation rather than the service provider. PIA tools are one of the many layers that will eventually be needed to protect user privacy.

### B. *Assessment of Cloud design*

Privacy designs need to be assessed at different phases of the design like in the initiation, planning, execution, closure and decommission phases [4]. The ignition phase should deal with the setting of high level privacy requirement recommendations, strategy and goals.

The Planning phase would elaborate on these requirements and goals and detail their inputs and outputs. The execution phase would be identifying the problems relating to the solutions which have been proposed and considering any alternative approaches if needed. Documenting issues and privacy exposures also are a part of this phase. In the closure phase, audits, change management processes, business continuity, disaster recovery are considered. Finally the decommission phase is to properly dispose the private and sensitive information obtained during the product's lifecycle. J.C. Cannon [11], [12] describes the processes and methodologies on how to integrate privacy considerations during development process. Auditing existing systems to identify privacy problem areas and protecting them against privacy intrusions is a competitive advantage for the product and the organization. At all phases, privacy experts need to be involved with adequate training.

### C. *Use of PETs*

Privacy Enhancing Technologies (PET) can be any technology that exists to protect or enhance an individual's privacy including facilitating individuals to their rights under various Acts and Laws [4]. Examples of such technologies include privacy management tools that enable inspection of server-side policies, secure access mechanisms for users to check and update their personal data, pseudonymization tools that allow users to withhold their true identity.

The Privacy Enhancing Technologies Symposium [13] addresses the aspects of privacy technologies, the design and realization of privacy services for the Internet, data systems and networks. This symposium brings together anonymity and privacy experts from around the world to discuss advances and new perspectives around the privacy of user's personally identifiable information.

In conclusion, since the benefits of PETs is huge to organizations, many technologies are being developed and debated such as Wallets of multiple virtual identities, anonymous credentials, Negotiation and enforcement of data handling, etc.

### D. *PccP Model*

The PccP model as described by Rahman [8], prescribes a three layered architecture. The layers include the Consumer Layer, the address mapping Layer and the Privacy Preserving Layer. The Consumer layer consists of the users while the address mapping layer helps in mapping the user and an IP address from a pool such that the user's actual IP address is made obscure. The transformed IP address is then used while navigating in the Cloud. The Privacy Preserved Layer has a unique user Cloud Identity Generator to generate a unique user Identity thereby ensuring the privacy of the user. Rahman proposes algorithms to generate the Unique Service Dependent Identity and Privacy preserver Match Logic.

To conclude, Rahman's Model attempts to enhance the privacy of sensitive user information such as IP addresses based on IP masking and unique identity generation until the user is present on the Cloud. However, this model needs further evaluation as to the extent of anonymity needed for the user.

### E. *Accountability for Cloud Services- A4Cloud.*

Cloud service providers lack accountability frameworks making it difficult for users to understand, influence and determine how their SLAs will be honored. Cloud services allow enterprises to outsource their business to third parties. The complexity of the services provider's eco-system may not be visible to the data user or enterprise. The A4Cloud project helps to create solutions to support users in deciding and tracking how their data will be used by the Cloud service providers by combating risk analysis, policy enforcement, monitoring and compliance auditing for security, assurance and redress.

The A4Cloud [2], [16] project is an Integrating project (IP) launched in 2012 in the EU's 7th Framework Program (FP7) led by HP Labs with many European countries as partners. The A4Cloud aims to enable the Cloud service providers to give the data users appropriate control and transparency over how their data is used and allow them to make choices about how the Cloud service providers protect their data in the Cloud. A4Cloud also aims to monitor and check compliance against user's expectations, business policies and regulations and lastly implement accountability ethically and effectively. An interdisciplinary co-design approach is the cote to the A4Cloud by combining legal, regulatory, polices, business processes and technical measure into a framework for accountability.

A4Cloud aims to [4]:

- Enable Cloud service providers to give their users appropriate control and transparency over how their data is used.
- Enable users to make choices about how Cloud service providers may use and will protect data in the Cloud.
- Monitor and Check compliance with user expectations, business policies and regulations.

Implement accountability ethically and effectively.

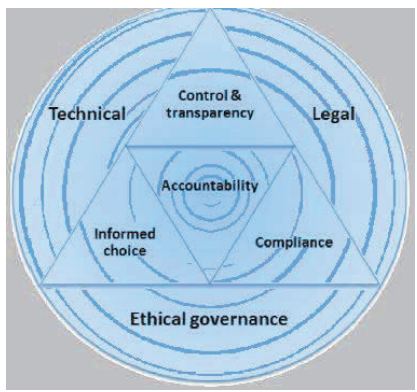

(Fig 1. A4Cloud interlocking Objectives [4])

A4Cloud Objectives: A4Cloud has four interlocking objectives to bring users, providers and regulators together in chains of accountability for data in the Cloud, clarifying liability and providing greater transparency overall.

Objective 1 [4]: Develop tools that enable Cloud service providers to give their users appropriate control and transparency over how their data is used, confidence that their data is handled according to their expectations and is protected in the Cloud, delivering increased levels of accountability to their customers.

Objective 2 [4]: Create tools that enable Cloud end users to make choices about how Cloud service providers may use and will protect data in the Cloud, and be better informed about the risks, consequences, and implementation of those choices.

Objective 3 [4]: Develop tools to monitor and check compliance with users' expectations, business policies and regulations. The A4Cloud provides a comprehensive accountability monitoring solution that would address the issue of preserving privacy and protecting confidential information.

Objective 4 [4]: Develop recommendations and guidelines for how to achieve accountability for the use of data by Cloud services, addressing commercial, legal, regulatory and end user concerns and ensuring that technical mechanisms work to support them.

In conclusion, A4Cloud Solution is a promising project that promises to address major barriers to trustworthy Cloud-based services. It helps to support service providers by using audited policy enforcement techniques, assessing and detecting policy violations, managing incidents and obtaining redress.

### F. *Legal as a Service (LaaS)*

Security, privacy and law-awareness are some of the biggest challenges faced by the Cloud service providers (CSP) to implement. Thus, Law-as-a-Service (LaaS) has been suggested for CSPs as a law-aware semantic Cloud policy infrastructure. The semantic legal policies in compliance with the laws are enforced automatically at the super-peer levels to enable LaaS. This allows CSPs to deploy their Cloud resources and services without worrying about law violations. Afterward, users could query data from the law-aware super-peer within a super-peer domain. Each query is also compliant with the laws. The law-aware super-peer is a unique guardian, who provides data integration and protection services for its peers within a super-peer domain. Each super-peer enforces the legal policies to enable data integration and protection services.

A privacy protection policy is a combination of ontologies and rules, where Description Logic (DL) based ontologies provide data integration, while Logic Program (LP)-based rules provide data query and protection services after data integration. Policies are shown as a combination of OWL-DL ontologies and stratified Datalog rules with negation for a policy's exceptions handling through defeasible reasoning. Law-as-a-Service (LaaS) enhances self-managed SaaS (System as a Service) on the automated security and privacy policy in the virtual data centers. Structure data is modeled as ontologies and used for data integration. Furthermore, the stratified Datalog rules with exceptions handling capabilities extend ontologies to enhance data protection and query services.

In summary, the concept of Law-as-a-Service (LaaS) [6] has been suggested by Hu, Wu and Cheng for CSPs as a law-aware semantic Cloud policy infrastructure. This seems as an exciting approach where a super-peer (unique law-aware guardian and trusted proxy) provides LaaS for its peers. The Super-peer also specifies how law compliant legal Cloud policies are enforced and unifies in the super-peer domain. This approach seems to be further developed and explored.

### IV. CONCLUSION

A search on online databases yielded very few published papers and articles on Legal challenges around Cloud Computing. This shows that this area of concern is not yet fully explored or discussed. A probable cause is due to the fact that Cloud boundaries are spread across geographies and each country has their own legal frameworks to deal with the Cyber world thereby complicating industry understanding of the Cloud and its legal complexities.

Cloud fears arise due to the perception of loss of control over sensitive data. The current control measures do not adequately address user's fears. Increased trust in the Cloud coupled with cryptographic techniques can help implement reliable controls thereby provide demonstrable business intelligence advantages to the Cloud stakeholders.

Cloud Providers are still fine-tuning their Service Level Agreements (SLAs) and Terms of Service (ToS) as the Cloud concept is yet in its infancy and unless users know the legal impact of Cloud Computing, the SLAs will not have the needed teeth to deal with legal issues arising out of the Cloud.

Additionally, laws need to be further enacted to deal with the Cloud designs. A broad international legal framework in cyberspace is the need of the hour as each country increases it's footprint in the Internet world. This kind of framework may best implemented by the United Nations to its member countries given its international reach.

## V. FUTURE WORK

The main issues related to cloud computing implementation are data-security, privacy, and law-awareness. By coupling legal compliance into CSP services, law-awareness can be incorporated into the cloud infrastructure. The concept of Law-as-a-Service (LaaS) [6] as suggested for CSPs is a law-aware semantic Cloud policy infrastructure. In this infrastructure framework, a super-peer (unique law-aware guardian and trusted proxy) provides LaaS for its peers. The Super-peer also specifies how law compliant legal Cloud policies are enforced and unifies in the super-peer domain. This approach seems to need further exploration and pilot-implementation.

(1) References

[1] Tom Kirkham, Django Armstrong, Karim Djemame, Marcelo Corrales, Mariam Kiran, Iheanyi Nwankwo, Ming Jiang, Nikolaus Forgo, "Assuring Data Privacy in Cloud Transformations," trustcom, pp.1063-1069, 2012 IEEE 11th International Conference on Trust, Security and Privacy in Computing and Communications, 2012, http://origin-www.computer.org.ezproxy.shsu.edu/csdl/proceedings/trustcom/2012/4745/00/4745b063.pdf.

[2] Siani Pearson, Vasilis Tountopoulos, Daniele Catteddu, Mario Sudholt, Refik Molva, Christoph Reich, Simone Fischer-Hubner, Christopher Millard, Volkmar Lotz, Martin Gilje Jaatun, Ronald Leenes, Chunming Rong, Javier Lopez, "Accountability for cloud and other future Internet services," cloudcom, pp.629-632, 4th IEEE International Conference on Cloud Computing Technology and Science Proceedings, 2012, http://origin-www.computer.org.ezproxy.shsu.edu/csdl/proceedings/cloudcom/2012/4511/00/06427512.pdf.

[3] George Kousiouris, George Vafiadis, Theodora Varvarigou, "A Front-end, Hadoop-based Data Management Service for Efficient Federated Clouds," cloudcom, pp.511-516, 2011 IEEE Third International Conference on Cloud Computing Technology and Science, 2011, http://origin-www.computer.org.ezproxy.shsu.edu/csdl/proceedings/cloudcom/2011/4622/00/4622a511.pdf.

[4] Siani Pearson, "Taking account of privacy when designing cloud computing services," icse-cloud, pp.44-52, 2009 ICSE Workshop on Software Engineering Challenges of Cloud Computing, 2009, http://origin-www.computer.org.ezproxy.shsu.edu/csdl/proceedings/icse-cloud/2009/3713/00/05071532.pdf.

[5] Masooda N. Bashir, Jay P. Kesan, Carol M Hayes, and Robert Zielinski. 2011. Privacy in the cloud: going beyond the contractarian paradigm. In *Proceedings of the 2011 Workshop on Governance of Technology, Information, and Policies* (GTIP '11). ACM, New York, NY, USA, 21-27.DOI=10.1145/2076496.2076499 http://doi.acm.org/10.1145/2076496.2076499.

[6] Yuh-Jong Hu, Win-Nan Wu, and Di-Rong Cheng. 2012. Towards law-aware semantic cloud policies with exceptions for data integration and protection. In *Proceedings of the 2nd International Conference on Web Intelligence, Mining and Semantics* (WIMS '12). ACM, New York, NY, USA, , Article 26 , 12 pages. DOI=10.1145/2254129.2254162 http://doi.acm.org/10.1145/2254129.2254162.

[7] Richard Chow, Philippe Golle, Markus Jakobsson, Elaine Shi, Jessica Staddon, Ryusuke Masuoka, and Jesus Molina. 2009. Controlling data in the cloud: outsourcing computation without outsourcing control. In *Proceedings of the 2009 ACM workshop on Cloud computing security* (CCSW '09). ACM, New York, NY, USA, 85-90. DOI=10.1145/1655008.1655020 http://doi.acm.org/10.1145/1655008.1655020.

[8] Syed Mujib Rahaman and Mohammad Farhatullah. 2012. A framework for preserving privacy in cloud computing with user service dependent identity. In *Proceedings of the International Conference on Advances in Computing, Communications and Informatics* (ICACCI '12). ACM, New York, NY, USA, 133-136. DOI=10.1145/2345396.2345419 http://doi.acm.org/10.1145/2345396.2345419.

[9] Electronic Privacy Information Center, "Cloud Computing", http://epic.org/privacy/cloudcomputing/.

[10] David Tancock, Siani Pearson, Andrew Charlsworth, University of Bristol, "A Privacy Impact Assessment tool for Cloud Computing", 2010, http://salsahpc.indiana.edu/CloudCom2010/slides/PDF/A%20Privacy%20Impact%20Assessment%20Tool%20For%20Cloud%20Computing.pdf.

[11] J.C. Cannon, "Review of Privacy: What Developers and IT Professionals Should Know", Addison Wesley, 2004, http://dl.acm.org.ezproxy.shsu.edu/citation.cfm?id=1071713.1071733&coll=DL&dl=ACM&CFID=326091868&CFTOKEN=96522449.

[12] Privacy Enhancing Technologies Symposium , "PETS 2013", http://petsymposium.org/2013/.

[13] Hein Timothy M Nguyen, Notre Dame Law School, 2012, "Cloud cover: Privacy protections and the stored communications Act in the age of cloud computing, http://www3.nd.edu/~ndlrev/archive_public/86ndlr5/Nguyen.pdf.

[14] Official Bill of Rights in the National Archives, "Bill of Rights", http://www.archives.gov/exhibits/charters/bill_of_rights_transcript.html .

[15] CGI, "Cloud Security for federal Agencies", 2012, http://www.cgi.com/files/white-papers/Cloud-Security-Federal_Agencies.pdf.

[16] Cloud Accountability Project (A4Cloud), http://www.a4cloud.eu/.

[17] C Ian Kyer, Gabriel M.A. Stern, Where in the World is My Data, "Jurisdictional issues with Cloud Computing", http://www.fasken.com/files/Event/3195cb2b-f29b-456d-8f98-7a3175930523/Presentation/EventAttachment/aea8833f-ab3c-48f1-854d-6ec4be989775/Jurisdictional_Issues_with_Cloud_Computing_Ian_Kyer_Gabriel_Stern.pdf.

[18] Lauren Gelman , The Center for Internet and Society, Ninth Circuit Court of Appeals: Stored Communications Act and Computer Fraud and Abuse Act Provide Cause of Action for Plaintiff, 2003, http://cyberlaw.stanford.edu/packets001500.shtml.

[19] From Title 18—CRIMES AND CRIMINAL PROCEDURE, 18 USC Ch. 121: STORED WIRE AND ELECTRONIC COMMUNICATIONS AND TRANSACTIONAL RECORDS ACCESS, [Online], http://uscode.house.gov/view.xhtml?path=/prelim@title18/part1/chapter121&edition=prelim